\documentstyle[prl,preprint,aps,epsf]{revtex}

\begin{document}
\tightenlines

\title{Nucleation in Systems with Elastic Forces}

\author{W.\ Klein,${}^{\S}$ ${}^{\dag}$ T.\ Lookman,${}^{\pm}$ A.\
Saxena,${}^{\pm}$ D.\ M.\ Hatch,${}^{\pm}$${}^{\star}$}

\address{${}^{\S}$CNLS, Los Alamos National Laboratory, Los Alamos,
NM 87545}

\address{${}^{\pm}$ T-11, Los Alamos National Laboratory, Los Alamos,
  NM 87545}

\address{${}^{\dag}$Department of Physics and Center for
Computational Science,\\ Boston University, Boston, MA 02215}

\address{${}^{\star}$Department of Physics and Astronomy, Brigham
  Young University, Provo, UT 84602}

\maketitle

\begin{abstract}

Systems with long-range interactions when quenched into a 
metastable state near the pseudo-spinodal exhibit nucleation
processes that are quite different from the classical nucleation seen
near the coexistence curve. In systems with long-range elastic forces
the description of the nucleation process can be quite subtle due to
the presence of bulk/interface elastic compatibility constraints. 
We analyze the nucleation process in a simple 2d model with
elastic forces and show that the nucleation process generates critical
droplets with a {\it different} structure than the stable phase. 
This has implications for nucleation in many crystal-crystal
transitions and the structure of the final state.

\end{abstract}
\bigskip
\noindent 
${}^{\dag}$ Permanent Address, ${}^{\star}$ Permanent Address



Nucleation in systems with long-range forces can be very different
\cite{ch,hk,mk,kl,so} than the process predicted by the classical theory\cite{lang,gunt}.
The reason for the difference is the presence of a
pseudo-spinodal(defined below) \cite{hks,nkr}
that affects the structure of the critical droplet\cite{hk,mk,kl} and
alters the dependence of the nucleation rate on the thermodynamic
parameters\cite{ch,uk}. An interesting and important class of materials
that exhibits pseudo-spinodal behavior is those that
interact through elastic forces; a subclass of which 
undergoes martensitic structural transitions.\cite{n}
An example of such a transition is when alloys such as FePd and NiTi
transform on cooling from an ``austenite'' phase at high temperatures to an
equal width mesoscale twin phase below the martensite transition 
temperature $T_{o}$\cite{o}.
This transition is first order and takes place via
nucleation. However, the nucleation process in these systems is not
well understood\cite{olson,reid,borg}. 

The purpose of this Letter is to present an analysis of nucleation
near the pseudo-spinodal of a model with elastic
forces. A complete specification of the critical droplet profile
requires, in addition to the usual bulk terms,\cite{mk,uk} a
consideration of a term generated by the elastic compatibility
constraints. This is, to our knowledge, the first treatment 
that describes nucleation near the pseudo-spinodal 
in this class of materials and is
the first indication that the compatibility constraint plays an
essential role in this nucleation process. In addition, this is the
first indication that homogeneous 
nucleation in a crystal-crystal phase transition
takes place with a critical droplet that does not have the symmetry of
the stable phase.

In this Letter we analyze a model that exhibits a 
two-dimensional 
square to rectangle transition\cite{slsb} and
captures the essential physics of crystal-crystal transitions in
systems with elastic forces.  
The order parameter(OP) is a rectangular or deviatoric
strain, which is a symmetry adapted combination of the 2d strain tensor
$\varepsilon_{\mu,\nu} (\mu,\nu = x,y)$. The non-OP, or secondary,
strain components are
related to the OP through a compatibility equation. We can
write a Ginzburg-Landau (GL) free energy in the form\cite{slsb}
$F(\epsilon) = F_{o}(\epsilon) + F_{grad}(\nabla\epsilon) +
F_{cs}(e_{1},e_{2})$, where
\begin{equation}
\label{free1}
F_{o}(\epsilon) = \int d{\vec r}\bigl\lbrack (\tau -
1)\epsilon^{2}({\vec r}) +
\epsilon^{2}({\vec r})(\epsilon^{2}({\vec r}) - 1)^{2}\bigr\rbrack,
\end{equation}

\begin{equation}
\label{free2}
F_{grad}(\nabla\epsilon) = \int d{\vec r}\bigl\lbrack {a\over
  4}(\nabla\epsilon({\vec r}))^{2} + {b\over
  8}(\nabla^{2}\epsilon({\vec r}))^{2}\bigr\rbrack,
\end{equation}

\begin{equation}
\label{free3}
F_{cs}(e_{1},e_{2}) = {1\over 2}\int d{\vec r}\bigl\lbrack
A_{1}e^{2}_{1}({\vec r}) + A_{2}e^{2}_{2}({\vec r})\bigr\rbrack.
\end{equation}

Here $\epsilon({\vec r}) = {1\over {\sqrt
    2}}\bigl\lbrack\varepsilon_{(xx)}({\vec r}) - \varepsilon_{(yy)}({\vec
    r})\bigr\rbrack$ is the OP strain, 
$e_{1}({\vec r}) = {1\over {\sqrt
    2}}\bigl\lbrack\varepsilon_{(xx)}({\vec r}) + \varepsilon_{(yy)}({\vec
    r})\bigr\rbrack$, $e_{2}({\vec r}) = \varepsilon_{(xy)}({\vec r})$
are the compression and shear strain, respectively
and $\tau = {(T - T_{c})\over (T_{o} - T_{c})}$, where $T_{c}$ is the
    temperature at which the OP would completely soften.
The subscripts $(xx)$, $(xy)$ and $(yy)$ denote partial derivatives,
$A_{1}$ and $A_{2}$ are elastic constants for the 
    compression and shear, and $a$ and $b$ are strain gradient
constants independant of $T$.

The St. Venant compatibility equation for the symmetric strain tensor is
$\nabla\times(\nabla\times\varepsilon({\vec r}))^{T} = 0$. Using the
Lagrangian multiplier formalism\cite{ka} in $d = 2$ we find for
Fourier expandable strains $\epsilon({\vec k})$ that $e_{1}({\vec k})$
and $e_{2}({\vec k})$ are proportional to $\epsilon({\vec k})$, the 
Fourier transform of $\epsilon({\vec r})$ with
${\vec k}$ dependent coefficients. This result allows us to replace
$F_{cs}(e_{1},e_{2})$ in Eq.(\ref{free3}) with an OP potential
$F_{cs}(\epsilon) = F^{bulk}_{cs}(\epsilon) +
F^{surface}_{cs}(\epsilon)$ where\cite{slsb,slsb2}
\begin{equation}
\label{free4}
F^{bulk}_{cs}(\epsilon) = \int d{\vec k}U^{bulk}({\vec
  k})|\epsilon({\vec k})|^{2},
\end{equation}.

\begin{equation}
\label{free5}
U^{bulk}({\vec k}) = {{4\rho\bigl\lbrack{(k^{2}_{x} -
  k^{2}_{y})\over k^{2}}\bigr\rbrack^{2}}\over 1 + 16\rho{k^{2}_{x}
  k^{2}_{y}\over k^{4}}},
\end{equation} 

\begin{equation}
\label{free6}
F^{surface}_{cs} = {A_{2}\over 2}\int d{\vec k}{|{1\over 2}\bigl\lbrack1 +
i{I({\vec k})\over J}\bigr\rbrack\epsilon({\vec k})|^{2}\over
|k_{y}|}\delta({\vec k} - {\vec k}_{surface}).
\end{equation}
where the surface term is generated by the interface between the high
symmetry ``austenite'' phase and the low symmetry ``martensite'' phase.
In Eq.(\ref{free6}) the integral is over the momentum modes of the interface,
$J$ is independent of ${\vec k}$ and depends only
on the ratio $\rho = A_{2}/2A_{1}$, and
\begin{equation}
\label{free7}
I({\vec k}) = {{k^{2}_{x} - k^{2}_{y}\over 2|k_{x}k_{y}|}\over 1 +
    \rho\bigl\lbrack {k^{2}_{x} + k^{2}_{y}\over 2|k_{x}k_{y}|}\bigr\rbrack^{2}}. 
\end{equation}
In the above equations $k_{x(y)}$ is the $x(y)$ component of ${\vec
  k}$ with respect to the underlying square lattice\cite{slsb2} 
and $k = |{\vec k}|$.

In the thermodynamic limit we can ignore $F^{surface}_{cs}$ and we
  first assume
  a spatially homogeneous $\epsilon({\vec r})$. The free energy has
  the following form; For $\tau>4/3$ there is one minimum at $\epsilon
  = 0$. For $1<\tau<4/3$ there are three minima but $\epsilon = 0$ is
  the global minimum. For $0<\tau<1$ there are again three minima. The
  one at $\epsilon = 0$ (the ``austenite'' phase) is metastable while
  the two minima, symmetric about $\epsilon = 0$ (the
  ``martensite'' phase) are stable. For $\tau<0$ there is no longer a
  minimum at $\epsilon = 0$. Note that this analysis does not consider
  instabilities to perturbations with non-zero wave vectors. We will
  return to this point below. We will first investigate   
the nucleation
  process from the metastable minimum at $\epsilon = 0$ near the
  spinodal at $\tau = 0$. From the spatially homogeneous free energy
  it is simple to calculate the order parameter exponent
  $\epsilon\sim\tau^{\beta}$ with $\beta = 1/2$. 
Reinstating the Laplacian term in free energy it can be seen that the correlation
  length diverges as $\xi\sim \tau^{-\nu}$ with $\nu=1/2$.
  Adding a spatially homogeneous external field to the free energy leads to a
  susceptibility $\chi\sim \tau^{-\gamma}$ with $\gamma = 1$\cite{sta}. 

Turning to nucleation near $\tau = 0$, we first
note that nucleation does not occur in systems with infinite range
interactions. That includes mean-field systems\cite{uk}. However,
systems with elastic forces do not have infinite range interactions
due to the screening from defects.\cite{eshel} Hence we will be
dealing with systems that have the same bulk and surface interactions
as in eqs.(\ref{free4}-\ref{free6}) except they will have an
exponential cutoff of the form $\exp(-{r\over R})$, where $R>>1$ is the interaction
range and $r = |{\vec r}|$.

Since the interaction range is large but finite the system is no
longer mean-field but near-mean-field\cite{bin,kle}. In order for the
mean-field approach, including the idea of a spinodal, to be a
reasonable approximation when $R\neq \infty$ the system must satisfy
the Ginzburg criterion\cite{bin,kle} namely
\begin{equation}
\label{gins}
{\xi^{d}\chi\over \xi^{2d}\epsilon^{2}} = {\tau^{-1}\over
  R^{d}\tau^{-{d\over 2}}\tau}<<1.
\end{equation}
Note that the correlation length,$\xi$, as
are all lengths, is in units of $R$. The Ginzburg criterion can be
rewritten as $R^{d}\tau^{2-d/2}= A>>1$ where $A$ is a fixed large number.
When the Ginzburg criterion is satisfied, many aspects of the
mean-field spinodal are still present. However, the singularity has
been smeared out.\cite{hks,nkr} The larger $A$ the better the spinodal
is approximated by the pseudo-spinodal. Since $A>>1$ for these systems
the pseudo-spinodal is very close to a true spinodal.

To calculate the nucleation or critical droplet structure we will use
saddle point techniques\cite{ch,kl,lang,uk}. Near the pseudo-spinodal there
will be an incompletely softened mode that can be identified by
examining the ${\vec k}$ coefficient of the Gaussian term in the
action. We take the action to be the free energy in 
Eqs.(\ref{free1}-\ref{free3})\cite{lang}.
Initially we will ignore the surface term given in Eq.(\ref{free6}).
The structure factor $S({\vec k})$ is then
\begin{equation}
\label{struct}
S({\vec k}) \sim \bigl\lbrack \tau + {\pm|a|\over 4}k^{2} + {b\over 8}k^{4}
+ U^{bulk}({\vec k})\bigr\rbrack^{-1},
\end{equation}
where the +(-) is for $a$ positive(negative). 
Consider first $a,b>0$. Since all terms in $S({\vec k})$ are positive
semi-definite the only divergence is when $k\rightarrow 0$ and $k_{x} = k_{y}$. The
surface term in Eq.(\ref{free6}) would appear to strongly suppress
the fluctuations that cause the divergence. 
However, this term was derived for a sharp
interface\cite{slsb,slsb2}. We need to extend this result to an
interface with width $\xi$. The reason for this particular scale will
become clear. We can consider the smooth interface to be a sequence
of sharp interfaces or steps each one contributing a term to the free energy of
the form given in Eq.(\ref{free6}). Since $k_{x} = k_{y}$, $I({\vec
  k}) = 0$. We will assume a form for the interface of 
$\epsilon_{interface}\propto \exp(-r/\xi)$ and $\xi>>R$.  
The
difference in order parameter amplitude between steps is 
\begin{equation}
\label{freed}
\Delta \epsilon_{interface} \propto {d\over dr}\exp(-r/\xi)dr =
-{\exp(-r/\xi)\over \xi}R,
\end{equation}
where we have chosen our differential step $dr = R$. This is justified
since all lengths are in units of $R$ and $R$ is the coarse
graining scale in the GL theory\cite{kl,uk}.  
Taking $k\sim
\xi^{-1}$ and $dk\sim \xi^{-1}$ in Eq.(\ref{free6})
the free energy cost for the surface is
$F^{surface} \sim {\xi^{2}\epsilon^{2}({\vec r})\over \xi}$,
where $\xi\epsilon({\vec r})$ scales as $\epsilon({\vec k})$ and the
number of steps in the surface is $\xi/R$. As we will see the dominant
contribution to the free energy $F$ scales as $\int d{\vec
r}\tau\epsilon^{2}({\vec r}) = \xi^{2}\tau\epsilon^{2}$, 
where by $\epsilon$ we mean to include only the dependence of
$\epsilon({\vec x})$ on $\tau$.
Comparing these two scaling forms we have 
$F^{surface} = F{C\over A^{1/2}}<<F$,
where $C$ is a constant of order 1 and $A>>1$ from the Ginzburg criterion. 
The Ginzburg criterion also implies that for a fixed but large $R$, $\tau$
cannot reach zero\cite{uk} and the system remain near-mean-field. 
The ratio $F^{surface}/F = 0$ only as $R$ and
hence $A\rightarrow \infty$. For the long-range potential we are using
the surface term is, in general, small enough to neglect. Note that if $A$ is not
infinite then the surface term is added to $S({\vec k})$
eliminating the divergence at $k=0$. Since $A>>1$ the structure factor
can be extremely large and the true spinodal is well approximated. 
In calculating the surface
contribution we have assumed that there is a domain with non-zero
$\epsilon({\vec x})$ with a linear size of the correlation length
$\xi$ imbedded in the metastable $\epsilon({\vec x}) = 0$ phase.  
We now proceed to demonstrate the existence of this domain. 
First we note that $a$ and $b$ in Eq.(\ref{free2}) must have units of 
length to the second  and fourth powers respectively and hence are 
proportional to $R^{2}$ and $R^{4}$ as all lengths must be
proportional to R\cite{uk}.
The Euler-Lagrange equation for the critical droplet is obtained by
setting the functional derivative of $F(\epsilon)$ in 
Eqs.(\ref{free1}-\ref{free3})
equal to zero to obtain
\begin{equation}
\label{el}
-{a\over 2}\nabla^{2}\epsilon({\vec r}) + {b\over
 4}\nabla^{4}\epsilon({\vec r}) + 2\tau\epsilon({\vec r}) -
 4\epsilon^{3}({\vec r}) + 6\epsilon^{5}({\vec r}) + \int d{\vec
 r}^{\prime}{\tilde U}^{bulk}({\vec r},{\vec r}^{\prime})\epsilon({\vec
 r}^{\prime}) = 0
\end{equation}
where ${\tilde U({\vec r})}^{bulk}$ is $U^{bulk}({\vec r})$ multiplied
by the exponential  cutoff $\exp(-{r\over R})$. We now assume a
solution of the form
\begin{equation}
\label{solution}
\epsilon({\vec r}) = \sum_{n}c_{n}(\tau) \exp(i{\vec k}_{o,n}\cdot{\vec
  r})\psi({{\vec r}\over L}) = G(\tau,{\vec r})\psi({{\vec r}\over L}),
\end{equation}
where $L>>R$, ${\vec k}_{o,0}$ is the value of ${\vec k}_{o,n}$ at
which the mean-field structure factor (Eq.(\ref{struct})) diverges and 
$c_{0}(\tau)>>c_{n}(\tau)$ for $n\neq 0$. For $\tau\sim 0$ the $c_{n}$
for $n\neq 0$ can be neglected\cite{kle}. We are near a pronounced 
pseudo-spinodal so that
we expect the critical droplet to have an interior structure similar
to spinodal critical fluctuations\cite{uk}(See Eq.(\ref{struct}).). 
Since $a>0$ implies ${\vec k}_{o,0} = 0$,
with the assumed form for $\epsilon({\vec r})$ in   
Eq.(\ref{solution}) the Euler-Lagrange equation becomes
\begin{equation}
\label{crdrop}
-{a\over 2} \nabla^{2}\psi({{\vec r}\over L}) + {b\over
 4}\nabla^{4}\psi({{\vec r}\over L}) + 2\tau\psi({{\vec r}\over L}) -
 4\psi^{3}({{\vec r}\over L}) + 6\psi^{5}({{\vec r\over L}}) = 0
\end{equation} 
Since $k_{x} = k_{y}$
the term involving ${\tilde U}^{bulk}$ gives no contribution and the
$c_{n}(\tau)$ are chosen so that $G^{3}(\tau,{\vec
  r}) = G(\tau,{\vec r})$.

Since $\tau\sim 0$ the solution of Eq.(\ref{crdrop}) has the scaled form 
\begin{equation}
\label{soln}
\psi({\vec r}) = D\tau^{1/2}{\tilde \psi}({B{\vec r}\over \xi})
\end{equation}
where the $\nabla^{4}\psi({{\vec r}\over L})$ and $\psi^{5}({{\vec
    r}\over L})$ 
terms have been neglected since they are higher order in $\tau$,
  $B$ and $D$ are constants that can be determined from Eq.(\ref{crdrop})
and $L = \xi$.
This form of the solution is what we assumed when we calculated 
the contribution of the surface term in Eq.(\ref{free6}) for a smooth
    interface. Hence the omission of the
  surface term is justified self consistently as is the scaling of the
  bulk free energy used to compare with the surface contribution. 

The nucleation barrier, $\Delta F$, is calculated by inserting the
critical droplet solution, Eq.(\ref{solution}), into the free
energy\cite{lang}, Eqs.(\ref{free1}-\ref{free3}). It is straightforward to see that
$\Delta F\propto R^{d}\tau^{2 - d/2} = A$. Therefore if $A = \infty$,
the system is mean-field, rather than near-mean-field, and there is no
nucleation.  

Note that the saddle point object which is the nucleation droplet
shows no evidence of the twin stripes seen in the simulation of the
stable phase of this model\cite{slsb}. The critical droplets near the
pseudo-spinodal are unstable\cite{hk,uk} and differ from the metastable phase 
by an order of magnitude given by $\tau^{1/2}\sim 0$. Their initial
growth phase is a ''filling in'' or an increase in the order parameter
difference.\cite{hk,uk} The filled in droplet will have a sharp
interface and hence must have twinning\cite{slsb}. Therefore,
the symmetry breaking which results
in the twin stripes must appear in the growth phase. As we will see,
if $a$, the coefficient of the $(\nabla\epsilon({\vec r}))^{2}$ term,  
is negative the case is somewhat different. We treat this next.  

For $a<0$ and $b>0$ we take the minus sign in the the structure factor
in Eq.(\ref{struct}). Since $U^{bulk}({\vec k})$ is independent of $k$,
the value of $k$ where the structure factor diverges is
\begin{equation}
\label{zero}
k = \pm{1\over 2}\bigl\lbrack {|a|\over 4} \pm ({|a|^{2}\over 16} - {(\tau +
  \tau_{o})b\over 2})^{1/2}\bigr\rbrack^{1/2},
\end{equation}
where $0\leq \tau_{o} \leq 4\rho$ is a fixed value of $U^{bulk}({\vec
  k})$.  For $\tau > {|a|^{2}\over 8b} - \tau_{o}$ there is
  no divergence for real $k$ and hence no instability. Since the
  largest value of $\tau$ for which there is an instability is the
  spinodal then for $a < 0$, the spinodal is at $\tau_{s} = {|a|^{2}\over
  8b} > 0$. The structure factor will now diverge at a non-zero value of
  $k = k_{o}$ where $k_{o}$ is given by Eq.(\ref{zero}) with
  $\tau_{o} = 0$. Note that the additional instability generated by
  $a<0$ is at a value of $\tau$ greater than $\tau = 0$ expected from
a simple thermodynamics calculation. It is straightforward to
  calculate the exponents of the correlation length and the order
  parameter which have the same values as those at the $\tau = 0$
  spinodal for $a>0$. 

Turning to the nucleation problem for $a<0$ and initially ignoring the surface
term, the Euler-Lagrange equation has the form 
\begin{equation}
\label{el2}
\int d{\vec r}^{\prime}{\tilde S}^{-1}({\vec   r}^{\prime})\epsilon({\vec r} - 
{\vec r}^{\prime}) -4\epsilon^{3}({\vec r}) + 6\epsilon^{5}({\vec r})
= 0,
\end{equation}
where ${\tilde S}^{-1}({\vec r})$ is given by the Fourier transform of
the inverse of $S({\vec k})$ in Eq.(\ref{struct}) with $U^{bulk}({\vec
  k})$ replaced by ${\tilde U}^{bulk}({\vec k})$. We again assume a
solution of the form given in Eq.(\ref{solution}) where ${\vec k}_{o,0}$
is the vector at which the structure factor diverges with $a<0$. Since
$\tau_{o} = 0$ implies $k_{x} = k_{y}$ and $|{\vec k}_{o,0}|$ is given by
Eq.(\ref{zero}), ${\vec k}_{o,0}$ is specified. We now expand
$\psi({{\vec r} - {\vec r}^{\prime}\over L})$ in a gradient expansion
about ${{\vec r}\over L}$ to obtain 
\begin{equation}
\label{el3}
{-|a|\over 2}\gamma_{2}\nabla^{2}\psi({{\vec r}\over L}) + (\tau -
\tau_{s})\gamma_{0}\psi({{\vec r}\over L}) - 4\psi^{3}({{\vec r}\over
  L}) = 0,
\end{equation}
where, anticipating the scaling,  
higher order derivatives and higher powers of $\psi({{\vec
    r}\over L})$ have been neglected. The constants $\gamma_{0}$ and
$\gamma_{2}$ are $\int d{\vec r}\lbrack\exp(i{\vec k}_{o,0}\cdot {\vec
  r}) + \exp(-i{\vec k}_{o,0}\cdot {\vec r})\rbrack{\tilde S}^{-1}({\vec r})$ 
and $\int d{\vec r}r^{2}\lbrack\exp(i{\vec k}_{o,0}\cdot {\vec r}) +
\exp(-i{\vec k}_{o,0}\cdot {\vec r})\rbrack{\tilde S}^{-1}({\vec r})$,
respectively. As above, the $c_{n}$ have been chosen so that
$G^{3}(\tau,{\vec r}) = G(\tau,{\vec r})$. The solution of
Eq.(\ref{el3}) is of the form
\begin{equation}
\label{soln3}
\psi({{\vec r}\over L}) = A(\tau - \tau_{s})^{1/2}{\tilde
  \psi}({B{\vec r}\over \xi})
\end{equation}
justifying the omission of higher order terms.

The strain field of the critical droplet 
\begin{equation}
\label{strain}
\epsilon({\vec r}) \sim A(\tau -
\tau_{s})^{1/2}\bigl\lbrack\exp(i{\vec k}_{o,0}\cdot{\vec r}) +
\exp(-i{\vec k}_{o,0}\cdot{\vec r})\bigr\rbrack{\tilde \psi}({B{\vec
    r}\over \xi}),
\end{equation}
where we have neglected terms with $n\neq 0$, is {\it not} that of the
stable phase but {\it does} exhibit a spatial modulation of regions where
$\epsilon({\vec r})\neq 0$. Consequently, the stable phase structure, 
as in the case $a>0$ where there is no spatial variation in the
strain, must evolve during the
growth phase. Note that the solution, Eq.(\ref{soln3}), justifies the
omission of the surface term via an argument virtually identical to
the one given above.

We have calculated the first critical droplet structures for
nucleation from an ``austenite'' like phase to a twinned ``martensite'' like
phase near the pseudo-spinodal in a system with elastic forces. The
droplets do not have the stable phase structure as expected from
classical nucleation\cite{lang,gunt} and in the $a>0$ case exhibit no
spatial modulation. Droplets that do not have the stable phase
structure have been predicted in the nucleation of the crystal from
the melt\cite{kl,so} but this is the first indication of such a
droplet structure in a crystal-crystal transition.
It is also the first result that demonstrates the importance of the
compatibility constraints to the phase transition kinetics. 

It is important to note in systems with $R>>1$ that classical
nucleation is strongly suppressed. In the classical case the
nucleation rate is proportional to $\exp(-{R^{d}\sigma^{d}\over
  {\Delta f}^{d-1}})$ \cite{lang,gunt} where $\sigma$ is the surface
tension between the droplet and the surrounding metastable state and
$\Delta f$ is the free energy density difference between the stable
and metastable states. For classical nucleation, near the coexistence
curve, $\sigma\sim 1$ so that for $R>>1$ nucleation is severely
suppressed. In order to have nucleation in a reasonable time frame the
quench must bring the system close to the pseudo-spinodal where
$\sigma<<1$. Therefore, nucleation near the pseudo-spinodal will
dominate the phase transition process in realistic
experiments. Finally we note that the form of nucleation discussed in
this Letter allows the possibility of evolution into metastable
crystallites with symmetries different than the stable phase. 

\begin{acknowledgements}
This work was performed under the auspices of the DOE at the Los Alamos 
National Laboratory under grants DE-FG02-95ER14498, LDRD-DR-2001501
and W-7405-ENG-36. W.K. gratefully acknowledges the support and
hospitality of the CNLS at LANL. D.H. gratefully acknowledges the
support and hospitality of the T-11 Group at LANL.
\end{acknowledgements}

\end{document}